\documentclass{article}
\usepackage[numbered]{algo}
\usepackage[section]{algorithm}
\usepackage{algpseudocode}
\usepackage{enumitem, soul}
\usepackage{varwidth}
\usepackage{tasks}
\usepackage{amsmath}
\usepackage{amssymb}
\usepackage{mathrsfs}
\usepackage{lmodern}
\DeclareFontFamily{OMX}{lmex}{}
\DeclareFontShape{OMX}{lmex}{m}{n}{<-> lmex10}{}
\usepackage{bigints}
\usepackage{graphicx}
\usepackage{indentfirst,csquotes}
\usepackage{booktabs}
\usepackage{float} 
\usepackage{relsize}
\usepackage{caption}
\usepackage{alphalph}
\usepackage{subfig}
\usepackage{svg}
\usepackage{natbib}
\algdef{SE}[VARIABLES]{Variables}{EndVariables}
   {\algorithmicvariables}
   {\algorithmicend\ \algorithmicvariables}
\algnewcommand{\algorithmicvariables}{\textbf{global variables}}

\usepackage{authblk}

\usepackage{etoolbox} 
\patchcmd{\subequations}{{0}}{{-1}}{}{}       
\patchcmd{\subequations}{\alph}{.\arabic}{}{} 

\begin{document}
\title{{\textit{Forecasting high frequency order flow imbalance using Hawkes processes}}}

\author[1]{Aditya Nittur Anantha}
\author[2]{Shashi Jain}
\affil[1] {\textsmaller[2]{SigmaQuant Technologies Pvt. Ltd., Indian Institute of Science}}
\affil[2] {\textsmaller[2]{Indian Institute of Science}}

\maketitle

\begin{abstract}
	Market information events are generated intermittently and disseminated at high speeds in real-time. Market participants consume this high-frequency data to build limit order books, representing the current bids and offers for a given asset. The arrival processes, or the order flow of bid and offer events, are asymmetric and possibly dependent on each other. The quantum and direction of this asymmetry are often associated with the direction of the traded price movement.  The Order Flow Imbalance (OFI) is an indicator commonly used to estimate this asymmetry. This paper uses Hawkes processes to estimate the OFI while accounting for the lagged dependence in the order flow between bids and offers. Secondly, we develop a method to forecast the near-term distribution of the OFI, which can then be used to compare models for forecasting OFI.  Thirdly, we propose a method to compare the forecasts of OFI for an arbitrarily large number of models. We apply the approach developed to tick data from the National Stock Exchange and observe that the Hawkes process modeled with a Sum of Exponential's kernel gives the best forecast among all competing models.
\end{abstract}
\section{Introduction}
Capital markets offer buyers and sellers a transparent and efficient mechanism to exchange goods for capital. In most modern capital markets, a continuous double auction mechanism is used where both the goods being offered for sale by a seller at a price (called the `Ask' price) and the price that a buyer is willing to pay (the `Bid' price) is continuously matched. Whenever the best ask price (the minimum price a seller is willing to accept) matches the best bid price (the maximum price a buyer is willing to pay), a trade is said to have occurred, and the goods offered by the seller are exchanged for the capital bid by the buyer. If a buyer or seller chooses, she can modify or cancel a current bid or offer. The buyer or seller (market participants) can choose to place a limit order or a market order; a limit order contains both the price and quantity the market participant wants to trade, while a market order contains only the quantity to be traded. The market disseminates the information of new bid and ask orders, modifications, cancellations, and subsequent trades as events (or ticks) on a tick-by-tick (TBT) channel. It is up to the buyers and sellers who participate or intend to participate in the market to listen to the TBT channel and construct the limit order book at their end. Market orders are not shown in the limit order book because they are transient events that do not persist in the market. They are traded immediately at the best available price. An example limit order book is shown below.\par
\begin{table}[H]
    \centering
    \begin{tabular}{|c|c|c|c|}
    \hline
        Bid Quantity & Bid Price & Ask Price & Ask Quantity \\ \hline
        100 & 99.05 & 100.15 & 120 \\ \hline
        110 & 98.75 & 100.25 & 100 \\ \hline
        140 & 98.65 & 100.35 & 90 \\ \hline
        110 & 98.55 & 100.45 & 60 \\ \hline
        60 & 98.45 & 100.55 & 10 \\ \hline
    \end{tabular}
\end{table}
The rules governing the elements of the limit order book, for example, the quantities that can be quoted, the minimum price change to improve on an existing Bid or Ask price, the minimum quantity that must be traded at a time, etc., are all market specific. These examples are a subset of all the rules a particular market might enforce, and the set of all such rules is called the market microstructure. \par
\subsection{Mechanism of trades and Order Flow Imbalance}
Trades can occur for one of two reasons: either a seller places an order such that the top buy price is matched, or a buyer places a new buy order such that the maximum sell price is matched. We distinguish between these two types of trades and call the former `SELL' trades and the latter `BUY' trades. If, over a time horizon, there are a more significant number of `SELL' trades compared to `BUY' trades, then there is selling pressure, and we can surmise that the price at which trades have occurred has either reduced or stayed the same and vice versa. The asymmetry in trading activity towards either the `SELL' side or `BUY' side is of interest to any market participant who may wish to revise her limit or market order price to maximize her chances of getting a trade at the lowest or highest price, depending on whether the market participant is a buyer or a seller. This bias in trading activity regarding trade events is termed the Order Flow Imbalance (OFI). \par
\subsection{Market participants and motivation for measuring OFI}
In the market micro-structure literature, market participants are classified based on the desired intent of trading. While these are not strict definitions, they provide a way of thinking about the trading activity disseminated by the market in the TBT channel and the market participant's motivation. Market participants are classified as market makers, arbitrageurs, and speculators. Market makers place limit orders on both sides of the limit order book. By doing so, they offer liquidity to the other market participants and provide a reference to start pricing their limit or market orders. Arbitrageurs trade the relative difference in the price of the same asset across markets or relatively price the same asset using multiple derivatives and trade away that difference. Speculators bet on the asset's future price and often trade on a single side of the limit order book. Each market participant is affected differently by the OFI; for instance, market makers are interested in avoiding adverse selection. Market participants of all kinds today trade algorithmically by writing computer programs that read the TBT data disseminated by the market and output limit or market orders. The amount of data in the TBT channel has exploded in recent years, with many contracts having more than 10 million ticks in a trading day. Due to the increased amount of data disseminated, some empirical artifacts observed in the TBT data require closer examination and modeling. An example of one such empirically observed artifact is the clustering of trades. It is observed in the TBT data that trade events tend to happen in clusters; that is, `BUY' trades in the past affect `BUY' trades in the present, and `BUY' trades in the past also affect `SELL' trades in the present and vice versa.\par 
 With the rise of algorithmic trading fueled by extensive Tick-By-Tick (TBT) data, empirical observations such as trade clustering have become apparent, prompting the need for OFI models capable of capturing cross-dependence between `BUY' and `SELL' trades over time. Adapting to these complexities is essential for accurately forecasting market dynamics and facilitating informed trading strategies within the continually evolving financial landscape. \par 

\subsection{Studies conducted on the effect of algorithmic trading on liquidity.}
According to \cite{Report_1}, as of 2015–16, algorithmic trading accounted for 40\% of trading activity in the cash market segment, 52.94\% in equity derivatives, and 37.8\% in currency derivatives on the National Stock Exchange. These figures highlight a significant presence of algorithmic trading in various segments and suggest the potential for further growth of such trading practices, especially in developing markets. \cite{2013a} reports that high-frequency algorithmic traders engaged in market-making, a subset of all algorithmic trading, contributed around	71.5\%	of the trading volume in August 2011 and 62.8\%	in February 2012 at NASDAQ-OMX Stockholm exchange. \par
\cite{2011b} studies the relationship between algorithmic trading (AT) and liquidity. They report a beneficial effect of AT on liquidity, noting that AT increases the realized spreads for market makers, indicating that market makers employ algorithms to trade both sides of the book. An earlier study, \cite{2005b}, studies a cross-section of such algorithms from different providers. Adverse selection for market makers, when using AT, can quickly build up costs when liquidity vanishes or suddenly decreases on one side of the limit order book. OFI is the net difference between the number of buy and sell market orders in a specified time window. The relationship between OFI and price movement has been studied in the literature by \cite{2014a}. \cite{2012a} explore the relationship between OFI of the market order count and the traded price during the May 6th, 2010 flash crash, with market orders defined as any order that results in a trade. The OFI of market orders, as defined in this paper, could be used as an estimate of adverse selection post facto, and the forecasted distribution of OFI can be used as an indicator of adverse selection in the future. In light of the findings of these studies, we believe a near-term forecast of OFI distribution to be helpful to the regulator in terms of surveillance and to the market maker in terms of enhanced risk management.\par
To model the order flow in high frequency, we need to include all real-time trade events to avoid information loss due to aggregation. For instance, a model using aggregated buy orders in a window for forecasting OFI can miss out on the clustering of order arrivals.   We model the order flow as a counting process with self and cross-excitation using the Hawkes process. The self and cross-exciting property of the point process model we have adopted helps capture the effect of clustering in trade events and can easily be extended to include more types of events on the limit order book.

\subsection{Studies using counting processes to model artifacts of liquidity.}
Trading volume as a counting process has been studied extensively in the literature related to market microstructure. In \cite{1999a}, trading volume is shown to be negatively associated with future price volatility. Increased trading volume as a response to the arrival of public information in the U.S. Treasury market results in lower price volatility in the future relative to the time of arrival of public information. OFI is one of the ways to estimate the asymmetry in demand and supply of a given asset. OFI is the normalized difference between the `BUY' and `SELL' trading volumes. The relationship between OFI and price formation is a well-studied topic related to high-frequency market micro-structure, especially in market crashes, with most empirical studies focusing on extreme market movements. Earlier studies of market crashes and OFI include \cite{1993a} and \cite{1989a} around the October 1987 crash. \cite{2002a} provide an OFI cross-sectional analysis of NYSE stocks for ten years. However, all of the studies mentioned use daily estimates of OFI and estimation methods for computing the OFI. \par
\cite{1992a} introduced the Probability of Informed Trading (PIN). 
\begin{equation}
    \text{PIN} = \frac{\lambda_i}{\lambda_i + \lambda_u}
\end{equation}

Where, $\text{PIN}$ represents the probability of informed trading, $\lambda_i$ is the arrival rate of informed trades and $\lambda_u$ is the arrival rate of uninformed trades.

In this formulation, the PIN represents the proportion of informed trading activity relative to the total trading activity in the market. It quantifies the likelihood that a given trade is initiated by an informed trader rather than an uninformed trader.\par
By examining order flow imbalances and transaction attributes, PIN offers insights into the behavior of informed traders in financial markets. PIN relies on the observation that informed traders possess private information about the true value of an underlying asset, influencing their trading decisions. Consequently, their trading activity disrupts the expected pattern of order flow, leading to deviations from the random order flow expected without private information. PIN aims to quantify this deviation and infer the presence of informed trading activity. The calculation of PIN typically involves analyzing order flow data over a specific time period, modeling the arrival of informed and uninformed trade events as Poisson processes, with the intensity of trading activity reflecting the interplay between informed and liquidity traders.\par
The PIN model's reliance on a constant rate imposes limitations when addressing trade clustering. To accommodate this, adjustments in window sizes for PIN computation are often made retrospectively, for example, in \cite{2012a}, which curtails its effectiveness in predicting clustered arrivals. Additionally, the assumption of independence between the arrival rates of informed and non-informed traders is challenging to justify in real-world markets. Instances of cross-excitation between `BUY' and `SELL' trades intermittently demonstrate the exchange of information among market participants, undermining the notion of independence. \par

\subsection{Contribution of this paper}
This paper uses Hawkes processes to build an indicator for the OFI, which captures the cross-dependence between `BUY' trades and `SELL' trades. We develop multiple forecasting models for the OFI and test the efficacy of each in relative terms. We develop a general framework for computing the loss function. This method can be used in conjunction with tests for superior predictive ability to identify a benchmark forecasting model among many competing models.\par 
Most studies in the literature use classification algorithms to determine the trade direction. \cite{2012a} uses a bulk trade classification algorithm; other studies have employed the Lee-Ready algorithm widely for trade classification. The Lee-Ready algorithm calculates trade direction by comparing the price of each trade to the preceding trade; trades occurring at prices above the previous trade are classified as buyer-initiated, while those below are considered seller-initiated. The bulk classification algorithm and the Lee-Ready algorithm may struggle in highly volatile markets or during periods of rapid price movements, potentially leading to incorrect classification. As shown in \cite{2018a}, high-frequency estimates of liquidity need to account for the increased short-term volatility in sub-second timeframes.\par 
Unlike the other studies mentioned earlier, we present a method of computing the OFI at high frequency (event or tick time) for market orders without employing any classification algorithm. The OFI used in this paper is not an estimate but the realized OFI with the buy/sell classification of the trade computed for each trade tick. Post classification of trade events, we follow the work of \cite{2012a} and start by modeling the market order arrivals as a counting process. We introduce a method by which OFI can be updated at any frequency, including tick time. To use the OFI as a valuable indicator for high-frequency market-making strategies, it becomes essential, especially given the increasing proportion of algorithmic trading, to provide a near real-time forecast of OFI, which could estimate, among other things, the near-term volatility, as shown by \cite{2022a}. Due to the noisy nature of high-frequency limit order book dynamics, as explored by \cite{2018a}, it is further necessary to forecast not just the expected near-term OFI but also the near-term distribution of the OFI. In this paper, we present a method to forecast a near-term distribution of the OFI.\par
Liquidity, in general, is not a precisely defined mathematical concept, making its study somewhat subjective. However, the bid-ask spread, order book volume, and the time distribution of market events are widely used to provide structure for analyzing liquidity in markets. In our work, we focus on the time distribution of market events, specifically on modeling the time distribution of market order events. We examine the effect of inter-trade arrival time on the dependency structure of the limit order book for market orders, in terms of the temporal dependence between the `BUY' and `SELL' market order flows. Our results show that adequately modeling the inter-arrival time of trades and the dependency structure of the limit order book, as detailed previously, leads to a significant improvement in forecast quality. \par
We structure the paper as follows --- we start by discussing the tick-by-tick data we have used and define OFI. Secondly, we describe the problem and models for forecasting the OFI. Thirdly, we define the problem statement, the methodology used for comparing models, and the methodology used for forecasting. We end the paper by presenting our results and the subsequent conclusions.
 
\section{Data}
We use processed tick data recorded on September 19th 2018 from the National Stock Exchange of Nifty futures expiring on September 27th 2018 (current month futures contract). We record ticks off the wire and filter for ticks of type `NEW\_TICK' (new tick), `MODIFY\_TICK' (modify tick), `CANCEL\_TICK' (cancel tick) and `TRADE' (trade tick). A snippet of the data used is provided in Table 1.
We mark a trade as either buy or sell by identifying whether a `BUY' or a `SELL' order resulted in the TRADE event. We use the two order IDs provided in the raw tick data for TRADE ticks.
\begin{table}[H]
	\tiny
		\centering 
    \caption{Ticks as received in NSE on September 2018 for NIFTY futures expiring 27th Sep 2018 }
	\resizebox{\textwidth}{!}{%
		\begin{tabular}{@{}llllllllllll@{}}
\toprule   Time              & Type & Symbol & Expiry & Event  & Side & Price    & Qty & Oid1             & Oid2             \\ \midrule
09:15:00.077863519 & FUT            & NIFTY  & 20180927   & NEW\_TICK    & BUY  & 11348.85 & 750 & \textbf{1100000000000928} & -1               \\
09:15:00.078110149 & FUT            & NIFTY  & 20180927   & MODIFY\_TICK & BUY  & 11319.8  & 75  & 1100000000000724 & -1               \\
09:15:00.078405918 & FUT            & NIFTY  & 20180927   & MODIFY\_TICK & BUY  & 11340.15 & 75  & 1100000000000770 & -1               \\
09:15:00.078495133 & FUT            & NIFTY  & 20180927   & MODIFY\_TICK & BUY  & 11338.15 & 75  & 1100000000000769 & -1               \\
09:15:00.079233914 & FUT            & NIFTY  & 20180927   & NEW\_TICK    & SELL & 11417.0  & 75  & 1100000000000929 & -1               \\
09:15:00.079445682 & FUT            & NIFTY  & 20180927   & NEW\_TICK    & SELL & 11349.9  & 75  & 1100000000000930 & -1               \\
09:15:00.079855028 & FUT            & NIFTY  & 20180927   & NEW\_TICK    & BUY  & 11315.0  & 75  & 1100000000000931 & -1               \\
09:15:00.080119943 & FUT            & NIFTY  & 20180927   & NEW\_TICK    & SELL & 11380.0  & 150 & 1100000000000932 & -1               \\
09:15:00.080125861 & FUT            & NIFTY  & 20180927   & NEW\_TICK    & BUY  & 11260.0  & 150 & 1100000000000933 & -1               \\
09:15:00.081216269 & FUT            & NIFTY  & 20180927   & NEW\_TICK    & BUY  & 11340.0  & 75  & 1100000000000935 & -1               \\
09:15:00.082875605 & FUT            & NIFTY  & 20180927   & NEW\_TICK    & BUY  & 11029.0  & 450 & 1100000000000937 & -1               \\
09:15:00.083489061 & FUT            & NIFTY  & 20180927   & TRADE        & SELL & 11348.85 & 75  & \textbf{1100000000000928} & 1100000000000938
\end{tabular}
}
\end{table}
We express $N_{t^{'}}$ as the counting process of trades as 
\begin{equation} \label{counting_1}
	N_{t^{'}}^{X} = \sum_{t=0}^{t{'}} \delta_{t}^{X} 
\end{equation}
where $t$ is the time of any trade event, $t{'}$ the time of the last trade event, $X$ can take values of either `BUY' or `SELL' and $\delta_{t}$ is the Dirac measure.
\subsection{Order Flow Imbalance}
We mark market orders as `BUY' or `SELL' depending on whether a `BUY' limit order filled a limit order standing on the sell side of the limit order book or vice versa. Tick data from NSE provides a convenient way of classifying the trade direction without resorting to estimation procedures like the eponymous Lee-Ready algorithm described by \cite{1991a}. The trade message from NSE contains two exchange order numbers. Exchange order numbers uniquely identify limit orders. The second exchange order number is that of the limit order that caused the trade; we call this the \textit{aggressive} order, and the first exchange order number is that of the standing limit order whose price was matched by the aggressive order, we call this the \textit{passive} order. Since we have all the tick data for the day, it is possible to look back in time and find whether the passive order was a `SELL' order or a `BUY' order by using the first exchange order number in the trade message. This allows for a deterministic classification of every trade.  Order Flow Imbalance is formally defined as:
\begin{equation} \label{OFI_1}
	OFI(T,h) = \frac {\Delta N^s_{T-h,T} - \Delta N^b_{T-h,T}} {\Delta N^s_{T-h,T} + \Delta N^b_{T-h,T}}
\end{equation}

and is measurable at time $T$, such that,
\begin{enumerate}
        \item $T \in \{t_1=0, .... , t_i, .... t_N = T_F \}$ where $t_i - t_{i-1} = 1$ $ \forall$ $i$
	\item $h$ is the window length over which trades are counted.
	\item	$\Delta N^s_{T-h,T} = N^s_{T} - N^s_{T-h}$ is the number of `SELL' classified trades over the window length $h$ and $T$ is the present time. 
	\item	$\Delta N^b_{T-h,T} = N^b_{T} - N^b_{T-h}$ is the number of `BUY' classified trades over the window length $h$ and $T$ is the present time. 
\end{enumerate}

\begin{table}[H]
		\centering 
    \caption{Summary statistics for OFI on 19th September 2018 for NIFTY futures expiring 27th September 2018}
			\begin{tabular}{@{}lllllllll@{}}
				\toprule Count & Mean                 & StdDeviation                & Min                 & 25\%                 & 50\%                & 75\%                & Max                \\ \midrule
				315 & -0.076 & 0.323 & -0.779 & -0.311 & -0.111 & 0.149 & 0.765
		\end{tabular}
\end{table}
\begin{figure}[H]
	\begin{subfloat}
		\centering
		\includegraphics[width=8cm,height=8cm,keepaspectratio]{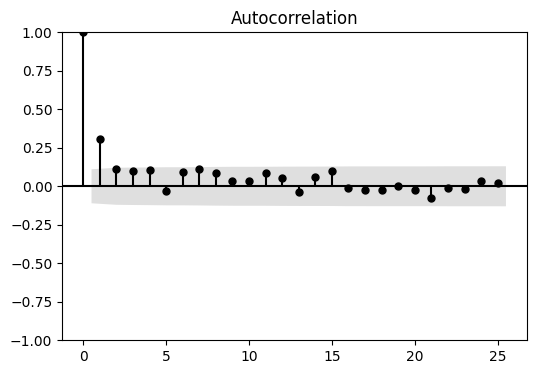}
	\end{subfloat}
	\begin{subfloat}
		\centering
		\includegraphics[width=8cm,height=8cm,keepaspectratio]{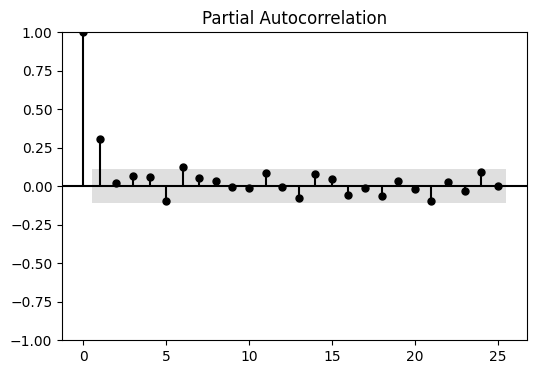}
	\end{subfloat}
	\caption{ACF and PACF of realized OFI}
\end{figure}
\begin{figure}[H]
	\begin{subfloat}
		\centering
		\includegraphics[width=8cm,height=8cm,keepaspectratio]{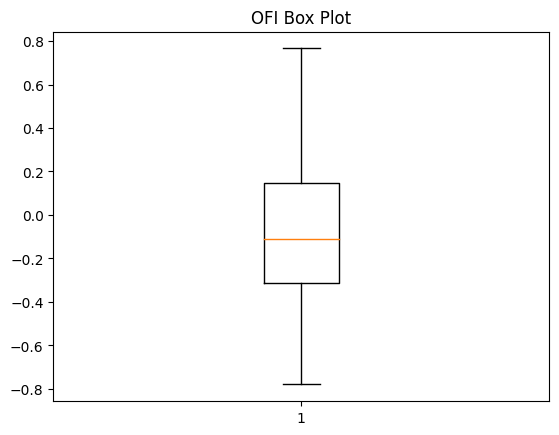}
	\end{subfloat}
	\begin{subfloat}
		\centering
		\includegraphics[width=8cm,height=8cm,keepaspectratio]{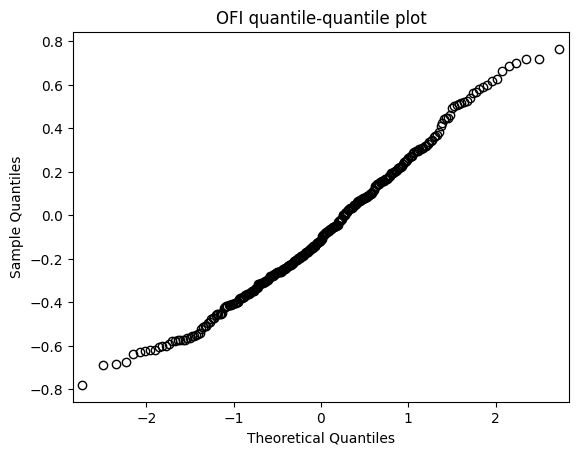}
	\end{subfloat}
\end{figure}
We construct a time series comprising minute-level OFI data for a single day, encompassing 375 observed OFI values. Subsequently, we perform the following tests to visually assess the presence of auto-regression within the time series.
The ACF and PACF both show significant contributions from autocorrelation, especially in the near past. This motivates the use of the Vector Auto Regression (VAR) model, among other models considered, which will be discussed later in this paper.

\section{Problem Description}
We aim to estimate E$[OFI(T+h, h) | \mathscr{F}_{T}]$, the expected value of OFI at time $T + h$, conditioned on the filtration $\mathscr{F}_{T}.$  We consider two classes of models. The first class jointly simulates the BUY (\(N^b_{T+h}\)) and SELL (\(N^s_{T+h}\)) market order arrival times, given \(\mathscr{F}_{T}\). These simulations are used to estimate the distribution of OFI for the window \((T, T+h]\). The second class constructs a buy and sell time series by aggregating the number of buy and sell market orders at regular intervals and models the expected OFI using linear time series forecasting techniques. \par
To forecast the OFI, we must incorporate the cross-sectional effects of buy-signed trades on sell-signed trades and vice versa in event time. We model the trade arrivals as Hawkes processes, which inherently model the cross effect. \cite{2015a} and \cite{2022b} model the market order point process using a Hawkes process. Still, there is little clarity on which Hawkes kernel is the better fit for modeling the trade arrival process and which Hawkes process is better suited for OFI forecasting. Note that accurately capturing the cross effect of the buy and the sell signed trades on each other is essential for the latter. \par 
To determine whether a benchmark kernel exists that outperforms all others, we provide a generic simulation-based algorithm that forecasts the near-term distribution of the OFI over a forecast window of 1 minute. To choose the best-performing model, we need a way to test the accuracy of the forecast across multiple models. 
The problem of comparing predictive accuracy is well known in the literature, starting with the Diebold-Mariano test (Diebold and Mariano, 1991), first proposed in a discussion paper for the Minneapolis Federal Reserve Bank. However, the Diebold-Mariano test is limited to comparing only two forecasts, and we need a way to compare multiple forecasts. White's reality check \citep{2000a} for data snooping provides a way to compare the forecast accuracy of multiple models by using a block bootstrapping method of random block size. We define the null hypothesis as in \cite{2000a} 
\begin{equation}
	\mathcal{H}_0: \max_{k=1,\ldots,m} \mathlarger{\mathop{\mathbb{E}}} [\Delta {L}_{k,t+1|t}(\Theta_k^{*})] \leq 0
\end{equation}
and the alternative hypothesis,
\begin{equation}
	\mathcal{H}_1: \max_{k=1,\ldots,m} \mathlarger{\mathop{\mathbb{E}}} [\Delta {L}_{k,t+1|t}(\Theta_k^{*})] > 0
\end{equation}
\raggedbottom
where $L$ is the loss function concerning a chosen realization or test set, $m$ the number of competing models, and $\Theta_k^{*}$ is the vector of estimated parameters and $\hat{\Theta}_k$ is the vector of true parameter values as computed from the data, such that $\Theta_k^{*} \xrightarrow{} \hat{\Theta}_k$, depending on the estimation method, for the k'\emph{th} competing model. $\mathcal{H}_0$ states that the benchmark model is not inferior to any of the $m$ competing models in terms of the difference in realized loss between the benchmark model and each competing model. Iteratively setting each $m$ competing model as the benchmark makes it possible to arrive at a single competing model or subset of competing models that outperform in terms of $\Delta L$. \cite{2005a} shows that while using the bootstrap procedure for constructing the null hypothesis in \cite{2000a}, the assumption under the null distribution, in practice, is $\mathlarger{\mathop{\mathbb{E}}} [\Delta {L}_{k,t+1|t}(\Theta_k^{*})] = 0$, additionally, it is shown that if $\max \mathlarger{\mathop{\mathbb{E}}} [\Delta \overline{L}]$ is negative, where $\Delta \overline{L}$ is the vector of sample averages of loss differences, then with probability one we get a degenerate distribution of the test statistic, when the benchmark model outperforms all other models, biasing the selection procedure against the benchmark model. \cite{2005a} provides a way to unbias the test statistic by normalizing and centering it.\par
We provide a method of comparing multiple modeling by defining an appropriate loss function and use the test for Superior Predictive Ability as defined in \cite{2005a} to find the best performing Hawkes kernel by fitting over a rolling time window of $W$ minutes and the forecasts are made for next $h$ minutes. \par
\section{Models for forecasting OFI}
\subsection{Hawkes Processes} 
Hawkes processes are a type of multivariate self-exciting counting process. They allow for the modeling of both self and cross-excitation between jointly distributed stochastic counting processes. In the case of OFI, there are two underlying counting processes --- one for those trade events that occur due to the arrival of `BUY' market orders and the second for those that arise due to the arrival of `SELL' market orders.
We follow the notation in \cite{2015a} to define a Hawkes process as a D-dimensional counting process, such that the stochastic intensity can be represented by:
\begin{equation} \label{hawkes_1}
	\lambda_t^i = \mu_i + \sum_{j=1}^{D} \bigintsss dN_{t^{'}}^j {\phi_{ij}}(t - {t^{'}})
\end{equation}
Where, $\lambda_t^i$ is the conditional stochastic intensity for the $i-th$ counting processs, and  ${N_{t^{'}}^j}$ is the $j$'th counting process defined as:
\begin{equation} \label{counting_2}
N_{t^{'}}^j = \sum_{t=0}^{t^{'}} \delta^j_t 
\end{equation}
where $\mu_i$ is an exogenous time invariant intensity and $\phi_{ij}(t)$ is such that:
\begin{enumerate}
	\item $\phi_{ij}(t) > 0$  $\forall$  $i,j$  where $1 \leq i,j \leq D$
	\item $\phi_{ij}(t) = 0$ if $t < 0$ $\forall$  $i,j$  where $1 \leq i,j \leq D$
	\item $\int \displaylimits_{0}^{\infty} \vert \phi_{ij}(t) \vert dt  < \infty$ $\forall$  $i,j$  where $1 \leq i,j \leq D$
\end{enumerate}
Using convolution notation,~\eqref{hawkes_1} can be represented as:
\begin{equation} \label{hawkes_2}
	 \Lambda_t = \Phi \ast dN_t 
\end{equation}
Where $\Lambda_t = {\{\lambda_i\}_{i=0}^{D}}$ is the vector of stochastic intensities, $\Phi = {\{\phi_{ij}\}_{i,j = 0}^{D}}$ is a matrix valued kernel and $N_t = \{N_t^{j}\}_{j=0}^{D}$ is the vector of counting processes. 
The kernel matrix $\Phi$ of the Hawkes process describes the relationship between the past and the present of the dimensions $D$ jointly. The Hawkes process allows to model the impact of the past timestamps of `BUY' signed trades on the future count of `BUY' and `SELL' signed trades (and vice versa), and therefore their impact on the distribution of the future OFI. These effects can be described as a finite number of parameters $\Theta$, resulting in a family of parametric kernels. $\Phi$ can also be expressed non-parametrically, resulting in a family of non-parametric kernels. This work uses parametric and non-parametric kernels to find the estimated Hawkes kernel $\hat{\Phi}$. \par
When~\eqref{hawkes_2} is used to model $\Lambda_{T,h}$ using $\Delta N^s_{T-h,T}$ and $\Delta N^b_{T-h,T}$, it takes the form:
\begin{equation}
	\begin{bmatrix}
		\lambda^s_{T-h,T} \\ 
		\lambda^b_{T-h,T}
	\end{bmatrix}
	= \Phi * \Delta N_{T-h,T}
\end{equation}
Where,
$
	\Delta N_{T-h,T} = 
	\begin{bmatrix}
		\Delta N^s_{T-h,T} \\
		\Delta N^b_{T-h,T}
	\end{bmatrix}	
$
and
$
	\Phi = 
	\begin{bmatrix}
		\phi_{ss} & \phi_{sb} \\
		\phi_{bs} & \phi_{bb}
	\end{bmatrix}
$.	
Given $\Delta N^s_{T-h,T}$ and $ \Delta N^b_{T-h,T}$ using the counting procedure described earlier, our first problem is to estimate the kernel matrix $\Phi$ to find $\hat{\lambda^s_{T}} = \lambda^s_{T-h,T}$ and $\hat{\lambda^b_{T}} = \lambda^b_{T-h,T}$. Our second problem is to forecast  $\Delta N_{t}^S$ and $\Delta N_{t}^B$ for $t \in (T, T+h]$, given the estimates $\hat{\lambda^s_{T-h,T}}$ and $\hat{\lambda^b_{T-h,T}}$. Using \eqref{OFI_1}, we get the forecasted $OFI$ for a given Hawkes kernel for a forecast window $h$. 

\subsection{Vector Auto Regression}
Vector auto-regression (VAR) is a statistical method for modeling the dynamic interdependencies among multiple time series variables. It is extensively used in various fields, including econometrics, finance, and macroeconomics. In a VAR model, each variable is regressed on its own lagged values and the lagged values of all other variables in the system. This allows for the examination of the simultaneous interactions among variables over time.

Formally, a VAR($p$) model with $k$ variables can be written as:
\[
\mathbf{y}_t = \mathbf{A}_1 \mathbf{y}_{t-1} + \mathbf{A}_2 \mathbf{y}_{t-2} + \cdots + \mathbf{A}_p \mathbf{y}_{t-p} + \mathbf{u}_t
\]
where $\mathbf{y}_t$ is a $k \times 1$ vector of time series variables, $\mathbf{A}_i$ are $k \times k$ coefficient matrices, and $\mathbf{u}_t$ is a $k \times 1$ vector of error terms. 
Applied to the problem at hand, we specify $y_t$ in the VAR model in terms of two dimensions as:  $\mathbf{y}_t = [\Delta N^b, \Delta N^s]^\top_t$, where $\Delta N^b_t$ and $\Delta N^s_t$ are the two time series of interest, the VAR model can be represented as follows. In a VAR($p$) model, each variable is a linear function of its own past values and the past values of the other variable. The model can be written as:

\[
\begin{bmatrix}
\Delta N^b\\
\Delta N^s
\end{bmatrix}_{t} = 
\mathbf{A}_1
\begin{bmatrix}
\Delta N^b\\
\Delta N^s
\end{bmatrix}_{t-1} + 
\mathbf{A}_2
\begin{bmatrix}
\Delta N^b\\
\Delta N^s
\end{bmatrix}_{t-2} + \cdots + 
\mathbf{A}_p
\begin{bmatrix}
\Delta N^b\\
\Delta N^s
\end{bmatrix}_{t-p} + 
\begin{bmatrix}
u^b \\
u^s
\end{bmatrix}_t
\forall t \in (T-h, T]
\]

Here, $\mathbf{A}_i$ are $2 \times 2$ coefficient matrices for $i = 1, 2, \ldots, p$, and $\begin{bmatrix} u^B \\ u^S \end{bmatrix}_t$ is a $2 \times 1$ vector of error terms. VAR models the relationship between $\Delta N^b_t$ and $\Delta N^s_t$, allowing for forecasting and analysis of how changes in the number of `BUY' market orders affects the `SELL' order flow and vice versa.\par

\section{Methodology}
\subsection{Fitting and Simulation}
\begin{figure}[H]
	\centering
	\includesvg[width=160mm]{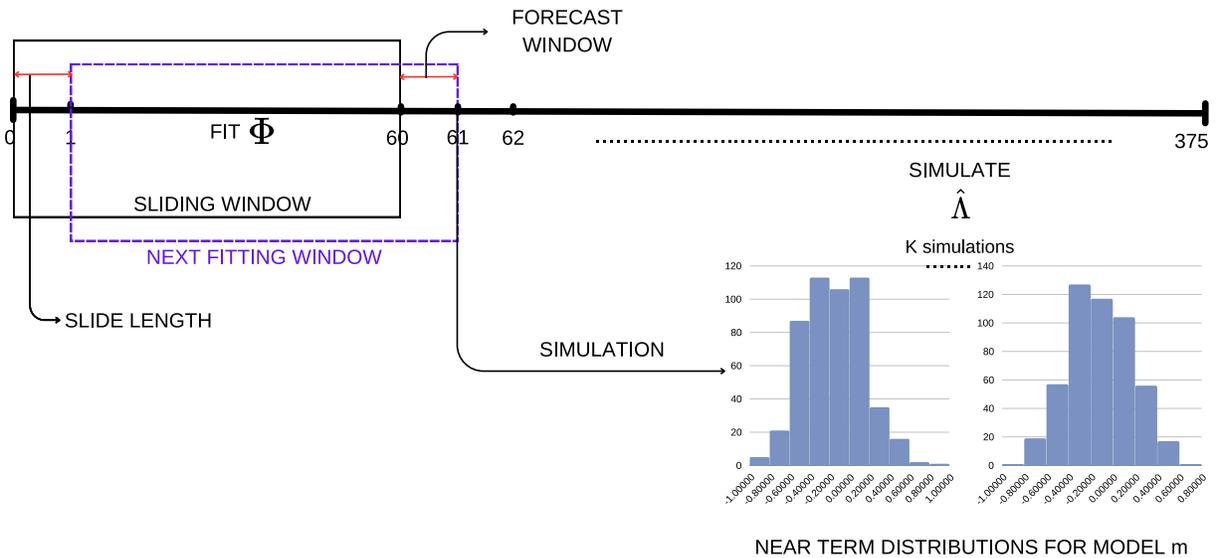}
	\caption{Model fitting and simulation}
\end{figure}

Using our established method, we begin by identifying each trade event and categorizing it as either `BUY' or `SELL'. Initially, we forecast arrays of timestamps by simulating $\hat{\Lambda}$. Where,
\begin{equation}
    \hat{\Lambda} = 
    \begin{bmatrix}
        \hat{\lambda}^b\\
        \hat{\lambda}^s
    \end{bmatrix}
\end{equation}
These timestamps, which correspond to the classified trade events, are then used to construct $\Delta N^b_{T-h,T}$ and $\Delta N^s_{T-h,T}$. Since $\hat{\Lambda}$ is already fitted on distinct trade events categorized by the trade side, the timestamp arrays come pre-classified as either `BUY' or `SELL'. The fitting window is set to 1 hour, and the forecast horizon is set to 1 minute. Therefore, for each 1-hour window, we forecast 1 minute of timestamps, from which we construct the forecasts of $\Delta N^b_{t}$ and $\Delta N^s_{t}$ for $t \in (T, T+h]$.\par

Next, we apply iterative modeling to this structured data. The chosen models are fitted to the data within each window. Once a model is fitted, we perform multiple simulations based on that model. Each simulation generates an array of arrival timestamps. The resulting $\Delta N^b_{t}$ and $\Delta N^s_{t}$, for $t \in (T, T+h]$,  from these simulations are used to calculate corresponding OFI values. These OFI values are collected to form arrays for each model. From these arrays, we construct an empirical distribution of OFI values, denoted as $\tilde{OFI}$, for each window. Each of these OFI distributions is then grouped by model into $\mathcal{F}_m$, as outlined in the general forecasting algorithm presented later in the paper.\par

After completing all iterations for a model $m$, the $\mathcal{F}_m$ array, which now contains the empirical OFI distributions for all windows, is appended to a larger set, $\mathcal{F}$. This process ensures that our forecasting and modeling methods are systematically applied and that the results are organized for further analysis and interpretation. \par

\subsection{Model comparison}
\begin{figure}[H]
	\centering
	\includesvg[width=160mm]{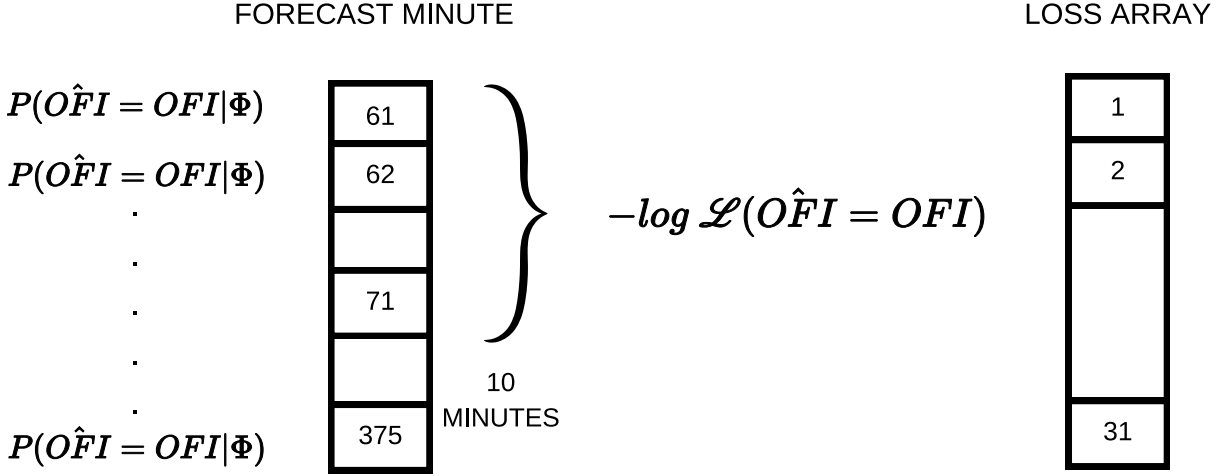}
	\caption{Computing the loss array}
\end{figure}
For each fitting window, we start by calculating the realized OFI. This involves assessing the difference between the volume of buy and sell orders within that specific window. Next, we determine the probability of this realized OFI occurring by comparing it to the empirical distribution $\tilde{OFI}$ that was constructed in the earlier stages of our analysis.\par

After determining these probabilities, we append them to a loss array, ensuring that there is one probability value for each window. This step is crucial as it allows us to evaluate the likelihood of the observed OFI within each time window for every model under consideration. By doing so, we can systematically assess the performance and reliability of our models across different time frames, leading to a comprehensive understanding of their behavior and accuracy.\par
\begin{figure}[H]
	\centering
	\includesvg[width=160mm]{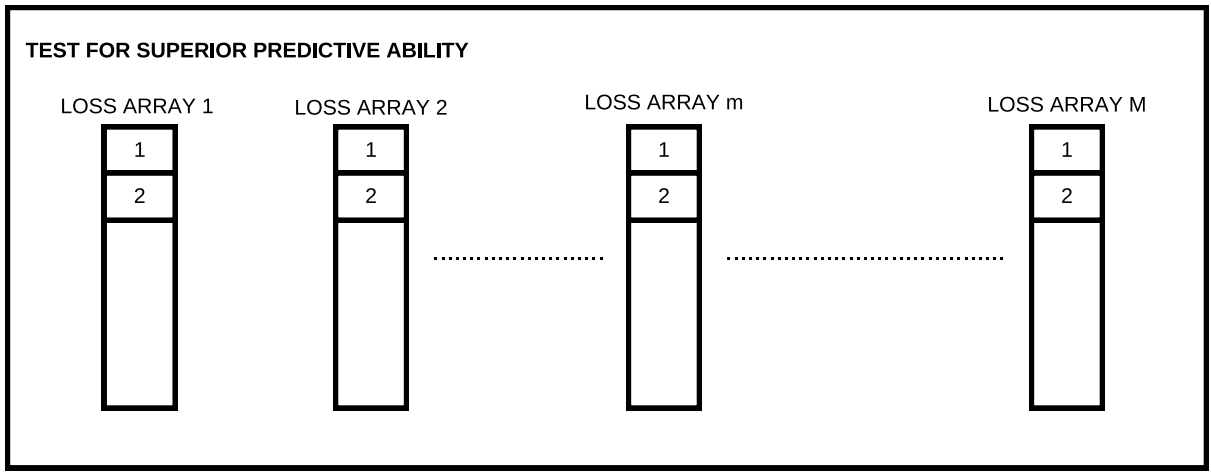}
	\caption{Model comparison method}
\end{figure}
Advancing the window by the forecast window length, we repeat this process, obtaining the probability of the present OFI value within each subsequent window. In our analysis, we present results based on a window length of one hour. The negative log-likelihood of these probabilities serves as our common loss function across models. We aggregate these probabilities and compute likelihood values over ten-minute intervals for each of the $M$ competing models. Finally, by comparing the negative log-likelihood arrays of the $M$ models under test, with each model acting as the benchmark, we assess superior predictive ability (SPA).

\subsection{Estimation of parametric Hawkes kernels}
In this section, we introduce a general method for estimating any parametric Hawkes kernel with an arbitrarily large number of parameters. A parametric Hawkes kernel can be represented in terms of its parameter vector as:
\begin{equation} \label{parametrization_1}
	\Phi(t) = f(\Theta, t)
\end{equation}
where $\Theta$ is the vector of parameters $[\theta_1, \theta_2, \ldots, \theta_n]$ \par
Rubin (1972a) shows for any point process, the joint density of occurrences is given by:
\begin{equation} \label{rubin_1}
	ln \mathcal{L}(\lambda(t, w_1)) = \bigintss_{0}^{T} ln(\lambda(t, w_1)) \cdot dN_t - \bigintss_{0}^{T} \lambda(t, w_1) \cdot dT
\end{equation}
where $\mathcal{L}$ is the joint density or likelihood function and $w_1$ is one realization of event times such that $w_1 \in [0,T]$. The expression for $\Lambda$ can be written in terms of the parameter vector as 
\begin{equation} \label{parametrization_2}
	\Lambda = g(\mu, \Theta, N_t, t)
\end{equation}
We use the method of maximum likelihood estimation in Ozaki (1977a) to find that $\hat{\Theta}$ which maximizes $ln \mathcal{L}$ or minimizes $-ln \mathcal{L}$. We use the stochastic gradient descent algorithm to find this minima. The stochastic gradient descent finds the minima of any convex surface by iteratively computing the gradient of the surface at each point and locally storing the value of the gradient. We express the gradient of the likelihood in terms of the parameter vector $\Theta$ as 
\begin{equation}
	\nabla_{\Theta} ln \mathcal{L}(\Lambda) = 
	\begin{bmatrix}
		\frac {\partial ln \mathcal{L}(\Lambda)} {\partial \theta_1} & \frac {\partial ln \mathcal{L}(\Lambda)} {\partial \theta_2} & \ldots & \frac {\partial ln \mathcal{L}(\Lambda)} {\partial \theta_n} 
	\end{bmatrix}	
\end{equation}
Once we have the expressions for $\nabla_{\Theta} ln \mathcal{L}(\Lambda) $ in terms of the parameters, we use the stochastic gradient descent algorithm as shown below to estimate that $\hat{\Theta}$ which minimizes $-\nabla_{\Theta} ln \mathcal{L}(\Lambda) $.
\begin{algorithm}[H]
	\begin{algo}{Stochastic Gradient Descent}{
	\label{alg:deep-learning-gradient-descent}
	\qinput{initial parameters $\Theta^{(0)}$, initial $\Delta \Theta_{min}$, initial $\Theta_{min}$, number of iterations $M$}
	\qoutput{final parameters $\Theta^{(M)}$}
}
	\begin{algorithmic}
	\Variables
		\State $\Delta \Theta_{min}$, global variable
		\State $\Theta_{min}$, global variable
		\EndVariables
	\end{algorithmic}\\	
		\qfor $i = 0$ \qto $M - 1$\\
			estimate $\nabla \mathcal{L}(\Theta^{(i)})$\\
			compute $\Delta \Theta^{(i)} = - \nabla \mathcal{L}(\Theta^{(i)})$\label{lin:deep-learning-delta-theta}\\
				select learning rate $\gamma$\\
			\qif $\Delta \Theta^{(i)} > \Delta \Theta_{min}$ \\ 
			\qthen \\
			\qreturn $\Theta_{min}$\\
			\qelse\\
				$\Delta \Theta_{min} = \Delta \Theta^{(i)}$ \\
				$\Theta_{min} = \Theta^{(i)}$ \\
				$\Theta^{(i + 1)} := \Theta^{(i)} + \gamma \Delta \Theta^{(i)}$\qfi\qrof\\
		\qreturn $\Theta^{(M)}$
	\end{algo}
	\caption[]{The stochastic gradient descent algorithm}
\end{algorithm}
The expression for the estimated $\Lambda$, $\hat{\Lambda}$ is written in terms of the estimated parameters $\hat{\Theta}$ using \eqref{parametrization_2} with $\Theta = \hat{\Theta}$. 
\subsection{Simulation of parametric Hawkes kernels}
The expression for $\hat{\Lambda}_t$,  $\hat{\Lambda}_t = g(u, \hat{\Theta}, N_t, t)$ is given as one of the inputs for generating simulated timestamps. We then use Ogata's modified thinning algorithm as presented in \citep{1981a}. 
\begin{algorithm}[H]
	\begin{algo}{Ogata's modified thinning algorithm}{
	\label{alg:hawkes-thinning-algorithm}
	\qinput{expression for $\hat{\Lambda}_t$, input parameters: time increment $\epsilon$, chosen such that $10^{-9} > \epsilon > 10^{-10}$ a small value and simulation time window length $\tau$}
	\qoutput{vector of simulated event times $\bar{T} = \begin{bmatrix}\bar{t}_1, \bar{t}_2, \ldots, \bar{t}_n \end{bmatrix}$}
	}
	\begin{algorithmic}
	\Variables
		\State array $\bar{T}$, global variable
		\State $t$, global variable
		\EndVariables
	\end{algorithmic}\\
		\qfor $t = 0$ \qto $\tau$\\
		  find new upper bound $M = \hat{\Lambda}_t(t+\epsilon)$ \\
			draw from exponential distribution $Exp(M)$: simulated time $t_M$\\
			draw from uniform distribution $Unif(0,M)$: value $u_m$\\
			\qif $u_M \leq \hat{\Lambda}_t(t_M)$ \\ 
			\qthen \\
				append $t$ to $\bar{T}$\\
			\qelse \\
				continue
			\qfi \qrof\\
			\qreturn $\bar{T}$
	\end{algo}
	\caption[]{Ogata's modified thinning algorithm}
\end{algorithm}
Ogata's modified thinning algorithm is used for simulating non-homogeneous Poisson processes. It improves upon the basic thinning algorithm by dynamically adjusting the intensity function over time, allowing for more accurate and computationally efficient simulations. The key innovation is the adaptive thinning mechanism, which reduces the number of potential events that need to be considered, thereby optimizing the simulation process. The algorithm is widely used in applications where the intensity function varies significantly, ensuring that the generated event times accurately reflect the underlying non-homogeneous process.\par
We evaluate the following Hawkes kernels for forecasting the OFI using the algorithms described in the earlier subsections. We have used the TICK library or the generic algorithm described above for fitting and simulation depending on the kernel. In order to study the effect of different methods of kernel simulation, we use two simulation methods -- the first implementation from the TICK library and the second implementation, Ogata's modified thinning algorithm. The parametric kernels which we use for forecasting the OFI are:
\begin{table}[H]
		\centering 
    \caption{Table of Hawkes parametric kernels}
\begin{tabular}{@{}llll@{}}
		\toprule
				 Hawkes kernel & Kernel expression & Method of estimation & Method of simulation  \\ \midrule
				 Exponential & $\phi(t) = \alpha e^{{-\beta}{t}} $ & TICK & TICK  \\ 
				 Sum of Exponential & $\phi_{i,j}(t) = \sum_{u=1}^{U}\alpha_{ij}^{u} e^{{-\beta^{u}}{t}}$ & TICK & TICK  \\ 
			 Power Law & $\phi(t) = \frac{\alpha}{{(\delta +t)}^{\beta}}$ & Modified SGD & Ogata's thinning algorithm  \\ 
\end{tabular}
\end{table}
In addition to evaluating the performance of parametric Hawkes kernels, we expand our analysis to include two non-parametric Hawkes kernels. The first of these non-parametric kernels is the Hawkes Conditional Law kernel, which is thoroughly described in the work by \cite{2014b}. The second non-parametric approach we incorporate is the expectation-maximization based Hawkes kernel estimation method, as detailed by \cite{2011a}. Including these non-parametric methods allows us to broaden our comparative analysis and provides a deeper understanding of different modeling techniques.\par

Alongside these kernel-based models, we also fit a VAR model. The VAR model is specifically employed to study the time evolution of OFI when it is viewed as a synchronous time series. This addition offers a distinct perspective on OFI analysis, contributing further depth to our study.\par

\begin{figure}[H]
	\centering
	\includesvg[width=\textwidth]{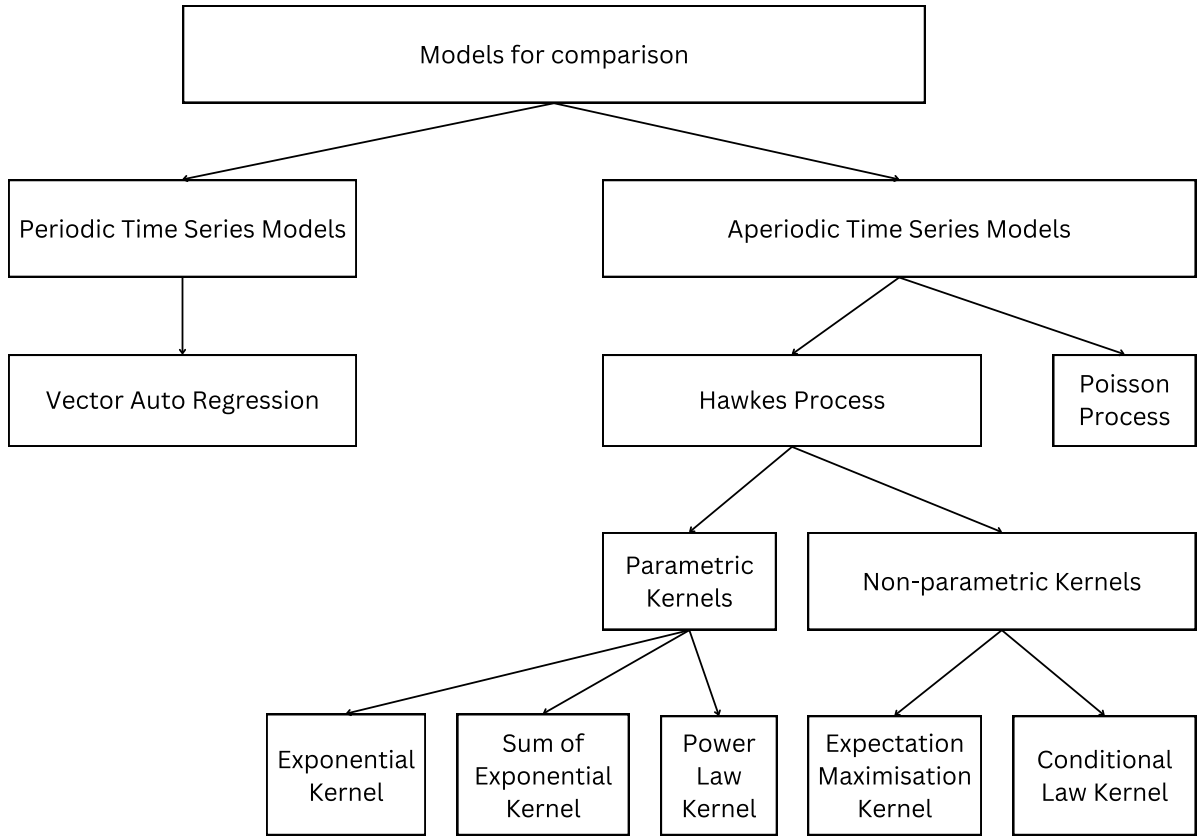}
	\caption{Compared models}
\end{figure}
\raggedbottom

Given the intrinsic differences between the parametric and non-parametric Hawkes models and the VAR model, the methods used to determine a commonly defined loss function vary across these different models. This variability in methodologies requires a detailed and clear explanation. Consequently, we dedicate the next subsection to thoroughly describing the forecasting methods developed for each model. In this subsection, we will provide a comprehensive explanation of how each model is handled, the specific steps involved in the forecasting process, and the rationale behind our comparative analysis.\par
  
\subsection{General forecasting algorithm}
\begin{algorithm}[H]
    \caption{Near term distribution $\Lambda$}
    \label{alg:forecasting}
    \begin{algorithmic}[1]
        \Procedure{NearTermDistribution}{$M, \{\Phi_m\}, W, \delta_W, N, K$}
            \State $\hat{\mathcal{F}} \gets [ ]$ \Comment{Global variable of near term distributions}
            \For{$m = 0$ \textbf{to} $M$}
                \State $\hat{\mathcal{F}}_m \gets [ ]$ \Comment{Near-term distributions for model $m$}
            \EndFor
            \While{$t \leq W$}
                \State Compute $T^B_i$ and $T^S_i$, the vectors of `BUY' and `SELL' arrays.
                \For{$m = 0$ \textbf{to} $M$}
                    \State Estimate $\Phi^i_m$.
                    \For{$s = 0$ \textbf{to} $K$}
                        \State Simulate $\hat{T^B_i}$ and $\hat{T^S_i}$.
                        \State Compute $\tilde{OFI}$ from $\hat{T^B_i}$ and $\hat{T^S_i}$.
                    \EndFor
                    \State Append $\tilde{OFI}$ to $\hat{\mathcal{F}}_m$.
                \EndFor
                \State Append $\hat{\mathcal{F}}_m$ to $\hat{\mathcal{F}}$.
            \EndWhile
            \State \textbf{return} vector of near term distributions $\mathcal{F}$ of size $M$.
        \EndProcedure
    \end{algorithmic}
\end{algorithm}

Tick-by-tick data recorded from the exchange is filtered for `TRADE' events. The events are classified as shown earlier. `BUY' and `SELL' timestamps arrays are the inputs for our algorithm. We divide the timestamps into one-minute windows. \par

The algorithm \textit{NearTermDistribution} is designed to compute near-term distributions for multiple models. The procedure takes as inputs the number of models \( M \), the set of parameters \( \{\Phi_m\} \) for each model, a window size \( W \), a step size \( \delta_W \), the number of simulations \( N \), and the number of iterations \( K \). Initially, it creates an empty global list \(\hat{\mathcal{F}}\) to store the near-term distributions for all models and initializes an empty list \(\hat{\mathcal{F}}_m\) for each model \( m \).

The algorithm proceeds by iterating through the time steps until the window length \( W \) is reached. For each time step, it computes the vectors \( T^B_i \) and \( T^S_i \), representing `BUY' and `SELL' orders, respectively. For each model \( m \), it estimates the model-specific parameters \(\Phi^i_m\) and performs \( K \) simulations to generate predicted `BUY' (\(\hat{T^B_i}\)) and `SELL' (\(\hat{T^S_i}\)) times. From these simulated times, it computes the near-term distribution of OFI \(\tilde{OFI}\) and appends this to the model's near-term distribution list \(\hat{\mathcal{F}}_m\). After completing the simulations for all models at the current time step, it appends \(\hat{\mathcal{F}}_m\) to the global list \(\hat{\mathcal{F}}\).

Once the loop finishes, the algorithm returns the vector of near-term distributions \(\mathcal{F}\) for all models, providing a comprehensive set of distributions for the specified models over the given time window.

\subsection{Forecasting using Hawkes processes}
We use the first hour of data for the initial forecast by concatenating the first 60 windows and fitting the Hawkes kernels in our test scheme. We use the fitted $\Phi$ to simulate one minute of trade arrivals. Since the Hawkes kernel fitted is two-dimensional, the forecasted arrivals contain a mix of `BUY' and `SELL' types. From the forecasted timestamps array, we compute the OFI. We conduct 500 such simulations, each resulting in an OFI value. We construct the empirical distribution (ED) of the forecasted OFI and the realized OFI for that one minute. Given the ED, we compute the probability of occurrence of the realized value of OFI. We now move the one-hour fitting window ahead by one minute, discard the first minute of data, incorporate the sixty-first minute, and repeat the process. The result is an array of 315 probabilities of occurrence of the realized OFI given the ED during the corresponding periods. We repeat this process for each Hawkes kernel and the Poisson process. We compute the negative log likelihoods from 10-minute windows of the 315 probability values across all models. Finally, we use the negative log-likelihood as the error function and evaluate the models' relative performance under the test for superior predictive ability. In Appendix IV, we have provided plots for the empirical distribution of the models under consideration and the realized OFI.\par.    
\subsection{Forecasting using Vector Auto Regression}
We investigate the performance of Hawkes models and VAR within the context of SPA, aiming to understand how modeling inter-arrival trade times affects forecast accuracy. In the VAR approach, we begin by computing $N^s_{T-h,T}$ and $N^b_{T-h,T}$ on a minute-by-minute basis, that is, with $h$ set to 1 minute, a departure from the previous method where the Hawkes process was fitted directly to trade times, thereby incorporating inter-arrival time information into the kernel $\Phi$. Now, we apply VAR to all 375-minute counts of `BUY' and `SELL' trade occurrences, capturing the dependency structure between these counts. Notably, the inter-order arrival time is excluded from the VAR model but retained in the Hawkes framework. Following the fitting of VAR to minute trade counts, we use the Gaussian error distributions of the model coefficients to simulate trade counts. The order of the VAR model, denoted as $p$, is determined using the Akaike Information Criteria. We fit the model on sliding windows of length $W$ minutes each and forecast counts for the subsequent minute, accounting for the respective error distribution of coefficients. Employing the previously outlined methodology, we sample from the Gaussian error distribution, incorporating the sampled error into the coefficients to forecast simulated counts. This process is augmented by introducing a loss array based on negative log-likelihood.\par

\section{Results}
\subsection{Test for Superior Predictive Ability --- parametric and non-parametric kernels}
        \begin{table}[H]
            \centering 
    \caption{Table comparing Hawkes parametric, non-parametric kernels and Poisson under the test for superior predictive ability}
		\begin{tabular}{@{}lllllll@{}}
            \toprule
		HawkesExp & HawkesSumExp & HawkesCondLaw & HawkesEM & Poisson & HawkesPowerLaw & VAR \\ \midrule
		0.002        & 0.743             & 0.257              & 0.0         & 0.0              & 0.0 & 0.101                 \\ \bottomrule
		\end{tabular}
	\end{table}
We iteratively use each of the $k$ models under consideration as the benchmark model and check the p-value under the test for superior predictive ability. A low p-value indicates that the benchmark model is inferior to at least one of the models under comparison. In contrast, a high p-value indicates that the sample data does not provide enough evidence to refute the null hypothesis. We conduct three tests: one for all the parametric models under consideration, one for all non-parametric models, and the last for both parametric and non-parametric models. We see that among the parametric models considered, the sum of the exponential Hawkes model is the best performing. Among the non-parametric kernels, the conditional law Hawkes model is the best performing. When we compare both parametric and non-parametric models together, we see that when either the sum of exponential or the conditional law Hawkes model is used, there isn't enough evidence presented by our data to reject the null that the two outperform the others. As a result, there isn't a clear-cut, better-performing model between the Hawkes sum of exponential and conditional law models. However, since the sum of the exponential model is far less computationally expensive than the conditional law, we favor the sum of exponential as the better choice. \par
Our findings indicate that during the test data period, the exponential parametric Hawkes kernel and the non-parametric Conditional Law Hawkes kernel, as proposed by Bacry and Muzy (2014), outperforms other methods in forecasting. Forecasting the rate of arrival of trade events proves to be computationally demanding due to the substantial volume of data generated in markets conducive to algorithmic trading. The process of fitting data to a model and forecasting necessitates a delta time exceeding the minimum inter-tick arrival time. The advantage of employing sum of exponential parametric models lies in their Markovian nature. This characteristic significantly reduces the time required for fitting, simulation, and forecasting as parameter estimation only relies on current data rather than historical data. Consequently, simulation becomes unnecessary, and the fitted parameters suffice for direct rate forecasting. This finding holds practical significance as it enables the utilization of the rate of arrivals in the immediately preceding time window for forecasting the arrival rate in the subsequent time window, thereby rendering the sum-of-exponential kernel computationally less burdensome and more suitable for risk management in high-frequency market-making strategies. It's important to note that when considering all the data within a day, the aforementioned results may exhibit a bias in favor of the sum-of-exponential Hawkes model. However, due to the nature of the SPA test, the Conditional Law and Vector Auto Regression models cannot be rejected with 95\% confidence. In our analysis, we employ the sum of exponential kernel Hawkes process to scrutinize the temporal dynamics of fitted Hawkes kernel norms within discrete one-hour time windows, further segmented into intervals ranging from 5 seconds to 5 minutes, enabling adjustment of granularity in our investigation. This approach aims to accurately capture the nuanced patterns of self-excitation and cross-excitation within the arrival time data. Subsequently, we visually track the evolution of $\phi_{i,j}$ values, indicative of interaction intensities between different event types, across the temporal spectrum.\par

\begin{figure}[H]
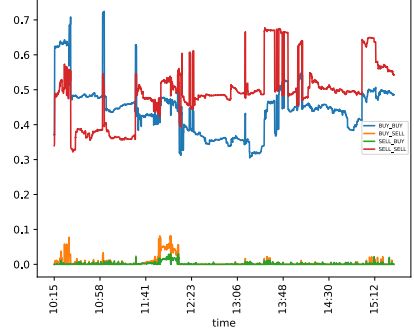

    \centering
    \subfloat[]{
        \includesvg[width=0.45\linewidth]{images/HawkesSumExp5m.svg}
    }
    \hfill
    \subfloat[]{
        \includesvg[width=0.45\linewidth]{images/HawkesSumExp1m.svg}
    }
    \vspace{0.4cm}
    \\
    \subfloat[]{
        \includesvg[width=0.45\linewidth]{images/HawkesSumExp30s.svg}
    }
    \hfill
    \subfloat[]{
        \includesvg[width=0.45\linewidth]{images/HawkesSumExp5s.svg}
    }
    \caption{NIFTY Futures expiring September 27th 2018}
\end{figure}

The five figures provide a comprehensive analysis of trade interactions using Hawkes sum of exponential kernels, offering insights into their dynamics. These interactions encompass a spectrum of scenarios, including the influence of past `BUY' trades on present `BUY' trades (BUY-BUY), the impact of past `BUY' trades on present `SELL' trades (BUY-SELL), the influence of past `SELL' trades on present `SELL' trades (SELL-SELL), and the effect of past `SELL' trades on present `BUY' trades (SELL-BUY). A nuanced understanding of market behavior emerges by examining the temporal evolution of these interactions. Across the trading day, BUY-BUY and SELL-SELL interactions dominate, indicative of sustained market sentiment. However, transient deviations manifest in heightened BUY-SELL and SELL-BUY interactions during intermittent intervals, revealing shifts in trading dynamics or brief episodes of intensified activity. This analysis underscores the multifaceted nature of market interactions between the buy and sell order flows, highlighting their evolving nature over time. 

\section{Conclusion}
Our study employs Hawkes processes and Vector Auto Regression to construct an Order Flow Imbalance indicator, capturing the interdependence between `BUY' and `SELL' trades. By advancing multiple forecasting models and establishing a general framework for computing the loss function, this study provides a robust method for identifying a benchmark forecasting model among various competing models. Unlike traditional classification algorithms, such as the Lee-Ready algorithm and bulk trade classification, which can falter in highly volatile markets, this paper presents a novel method for computing high-frequency OFI without relying on these classifications. The approach of using realized OFI with tick-by-tick trade classification allows for more precise and real-time updates, enhancing the utility of OFI as an indicator for high-frequency market-making strategies. Given the growing prevalence of algorithmic trading, the ability to forecast near-term OFI and its distribution is crucial for estimating market volatility and improving liquidity assessment. By modeling the time distribution of market order events and the temporal dependence between `BUY' and `SELL' market order flows, this paper demonstrates that incorporating the inter-arrival time of trades significantly enhances the confidence with which a benchmark model can be chosen in high-frequency settings. These findings underscore the importance of accurate real-time data analysis in understanding and predicting market behavior, ultimately aiding market participants in developing more informed trading strategies and effective risk management practices.\par
We observe a significant enhancement in forecast quality when considering the dependency structure of the limit order book through inter-trade arrival times. This paper highlights the potential of OFI as a valuable indicator for high-frequency trading strategies.

\bibliography{references}	
\section{Appendix-I}
\subsection{Test for stationarity of realized OFI}
\begin{table}[H]
		\centering 
    \caption{Augmented Dicky-Fuller test for stationarity}
		\begin{tabular}{@{}lllllll@{}}
                \toprule   ADFStatistic              & PValue & NoOfSamples & CriticalValues: &  1\%  & 5\% & 10\%             \\ \midrule
			-12.878 & 4.708e-24 & 314 &  & -3.451 & -2.870 & -2.571 
		\end{tabular}
\end{table}
\subsection{Tests for normality of realized OFI}
\begin{table}[H]
		\centering 
    \caption{Normality tests of OFI on 19th September 2018 for NIFTY futures expiring 27th September 2018}
			\begin{tabular}{@{}llll@{}}
				\toprule No. & Method                            & TestStatistic      & PValue                \\ \midrule
			1 & Kolmogorov-Smirnov normality test & 0.266 & 3.010e-20 \\
			2 & Shapiro-Wilk normality test       & 0.984 & 0.001 \\
			3 & Anderson-Darling normality test   & 1.090 & 0.007
	\end{tabular}
\end{table}
\section{Appendix-II}
\subsection{Test for Superior Predictive Ability --- parametric kernels}
\begin{table}[H]
		\centering 
    \caption{Table comparing Hawkes parametric kernels and Poisson under the test for superior predictive ability}
\begin{tabular}{@{}lllll@{}}
		\toprule
				HawkesExp & HawkesSumExp & HawkesPowerLaw & Poisson  \\ \midrule
				0.003        & 0.481             & 0.0                 & 0.0             
\end{tabular}
\end{table}
\subsection{Test for Superior Predictive Ability --- non-parametric kernels}
\begin{table}[H]
		\centering 
    \caption{Table comparing Hawkes non-parametric kernels and Poisson under the test for superior predictive ability}
\begin{tabular}{@{}llll@{}}
		\toprule
	 HawkesCondLaw & HawkesEM & Poisson \\ \midrule
	 0.518              & 0.0         & 0.0
\end{tabular}
\end{table}

\end{document}